\documentclass{ws-ijmpb}

\begin{document}

\markboth{K. Nakata}
{  Spin Currents  in  Insulators }

%
\catchline{}{}{}{}{}
%

\title{ TEMPERATURE DEPENDENCE OF SPIN CURRENTS  
     IN  ONE- AND THREE-DIMENSIONAL  INSULATORS
  }

\author{KOUKI NAKATA}

\address{Yukawa Institute for Theoretical Physics,\\ 
Kyoto University,\\ 
Kitashirakawa Oiwake-Cho, Kyoto 606-8502, Japan\\
nakata@yukawa.kyoto-u.ac.jp
 }

\maketitle

\begin{history}
\received{Day Month Year}
\revised{Day Month Year}
\end{history}

\begin{abstract}
The temperature dependence of spin currents in insulators at the finite temperature  near zero Kelvin  
is theoretically studied.
The spin currents are carried by Jordan-Wigner fermions and magnons   in one- and three- dimensional insulators.
The quasiparticle description of one-dimensional spin systems is valid 
only in the finite temperature near zero Kelvin.
These spin currents are generated by the external magnetic field gradient along the quantization axis
and also by the two-particle interaction gradient.
In one-dimensional insulators,
quantum fluctuations are strong  and 
the  spin current  carried by Jordan-Wigner fermions  shows the stronger dependence on temperatures
 than the one by magnons.
\end{abstract}


\section{Introduction}
\label{sec:intro}

Recently a new branch of physics and nanotechnology called 
spintronics\cite{maekawa,mod2} has emerged  and  
spin transport phenomena in condensed matter systems have been attracting special interests 
because of applications to spintronics.
This research field has seen a rapid development over the last decades.\cite{david}
The aim of spintronics is the control and utilization  of the spin as well as charge degrees of freedom of electrons.
Spintronics avoids the dissipation from Joule heating by replacing charge currents with spin currents,
and therefore  to clarify the properties of spin currents is an important theoretical issue
from viewpoints of fundamental science and potential applications.\cite{wolf,mod}

The spin current means a flow of the spin angular momentum, in general.
It also flows in insulators as well as in metals 
where conduction electrons carry  a spin current. 
In insulators there is no conduction electrons,
but there exists an other kind of carrier, namely, spin-waves.

Experimentally, 
a spin-wave spin current, a spin current carried by spin-waves
has already been established as a physical quantity.
Y.Kajiwara et al.\cite{spinwave}  have reported that 
a spin-wave spin current in an insulator can be generated and detected 
using direct and inverse spin-Hall effects.\cite{kato}
Moreover through the effects,
it is possible to convert an electric signal in a metal into a spin-wave spin current, and vice versa.
This spin-wave spin current has a novel feature that 
this current  persists for much greater distance than the conduction electron spin current, 
which disappears within very short distance (see Table \ref{tab1} ). 
This novel feature of the spin-wave spin current is expected to lead to developments beyond silicon-based technologies.

Theoretically,
Meier and Loss\cite{meier} have studied the magnon transport 
in both ferromagnetic and antiferromagnetic materials.
They have found that the spin conductance is quantized in the units of order $(g\mu_{\rm{B}})^2/h $
in the antiferromagnetic isotropic materials
($g$; the gyromagnetic ratio, $\mu_{\rm{B}}$; the Bohr magneton, $h$; Planck constant).
Fujimoto\cite{fujimoto} has studied a spin Hall effect of spin-waves in frustrated magnets.
A longitudinal magnetic field gradient induces a transverse spin current carried by spin wave;
$ J_x =  \sigma _{xy}^{\rm{SHE}}  \partial B/\partial y,   \sigma _{xy}^{\rm{SHE}}  \propto T^5 $ at the low-temperature,
where $ J_x  $ is the spin Hall current,
$  \sigma _{xy}^{\rm{SHE}}   $ is the Hall conductivity for the spin current and 
$ B $ is the magnetic field along the quantization axis.
Katsura et al.\cite{katsura,onose,matsumoto} have reported 
the intrinsic thermal Hall effect for magnons  due to the anomalous velocity.

Though thus theoretical studies of spin currents in insulators  have developed,
it should be emphasized that, besides magnons, 
there exists an other kind of carrier in insulators, namely, the Jordan-Wigner (J-W) fermion.
This carrier is peculiar to one-dimensional spin systems.
In order to clarify the  properties of spin currents carried by J-W fermions,
we focus on the one-dimensional spin-$1/2$ XXZ model.\cite{fradkin,chain,gia}
The  model reads
\begin{eqnarray}
 {\mathcal{H}}_{\rm{XXZ}} =   \Gamma    \sum_{i}    ( S_i^x S_{i+1}^x    +   S_i^y S_{i+1}^y  +   \Delta_{i} S_i^z S_{i+1}^z  ),
\label{eqn:z8}
\end{eqnarray}
where $ S_l^{\alpha } $ is $ \alpha  $ component of a spin-$1/2$ operator on $ l $ th site ($ \alpha =x,y,z $).
The constant parameter $ \Gamma  $  is the exchange coupling constant,
and $ \Delta   $ is the anisotropic interaction parameter.
This  model can be well mapped into spinless fermion systems via the J-W transformation.
When $ \Delta  = 1  $ the XXZ model reduces to the Heisenberg model,
and
when $ \Delta  = 0  $ this model reduces to the XY model;
$  {\mathcal{H}}_{\rm{XY}}   =   \Gamma \sum_{i} ( S_i^x S_{i+1}^x +   S_i^y S_{i+1}^y ) $,
which describes the disorder phase; $ \langle     S_l^{\alpha }    \rangle   =0  $.
After a transformation which respects the spin commutation relations;
 $ S_i^x \rightarrow   (-1)^i S_i^x, \ \ S_i^y \rightarrow   (-1)^i S_i^y, \ \   S_i^z  \rightarrow  S_i^z, $
the XXZ model changes the sign of  parameters as
  $  \Gamma  \rightarrow    - \Gamma, \ \ \Delta   \rightarrow   \Delta.  $ 
Thus regardless of the signs of the parameters,
it is enough to consider the case $ \Gamma  >0  $.
This paper  focuses on the properties of the spin current 
when   the value of $ \mid  \Delta   \mid   $ is smaller than one.

Quite recently, theoretical studies for spin currents in one-dimensional spin chains also have been rapidly progressing.  
Trauzettel et al.{\cite{loss2} have calculated the ac magnetization current and the power absorption of the XXZ model.
Hoogdalem et al.\cite{xxz5} have considered rectification effects\cite{rec1,rec2} through bosonization techniques
in both ferromagnetic and antiferromagnetic systems.

In this paper, we study the effects of external magnetic field gradients along the quantization axis
and two-particle interaction gradients on the spin current carried by J-W fermions in one-dimensional insulators;
the quasiparticle description of one-dimensional spin systems is valid 
in the finite temperature near zero Kelvin.\cite{fermiliquid,fujimoto1dim}
The magnetic field  is equivalent to a chemical potential for J-W fermions 
and the gradient plays a role similar to an electric field in electron systems.\cite{Haldane}
The two-particle interaction arises from the anisotropic interaction in eq.(\ref{eqn:z8})
and the gradient also generates the spin current.
The main aim of this paper is to reveal the temperature dependence
of the spin current carried by J-W fermions
and then compare it with that of the magnon current
in the finite temperature near zero Kelvin.

This paper is structured as follows.
In the one-dimensional XXZ model, the spin degrees of freedom can be 
mapped efficiently into fermion degrees of freedom via the J-W transformation.
Thus first,
we define microscopically the spin current density carried by J-W fermions. 
Second,
we evaluate it through the standard procedure of the Schwinger-Keldysh Green's function.
The effects of magnetic field gradients and two-particle interaction gradients
on the spin current are clarified in Sec.\ref{sec:intro}.
Then the magnon current generated by the two-particle interaction (i.e. magnon-magnon interaction)
is also calculated in Sec.\ref{sec:mag}.
Last, 
the temperature dependence of the spin current carried by Jordan-Wigner fermions is compared with 
that of the magnon current in Sec.\ref{sec:sc}. 
We discuss the cause for the stronger dependence  of the spin current carried by J-W fermions 
on temperatures than that of the magnon current in the finite temperature near zero Kelvin.

\begin{table}[ph]   
\tbl{The features  of spin currents; spin currents carried by conduction electrons (i.e. conduction electron spin currents),
by magnons (i.e. magnon currents) and by J-W fermions (i.e. J-W spin currents).
The magnon current, which is the quantized spin-wave spin current,
 persists for much greater distance than the conduction electron spin current
  because the dissipation from Joule heating does not exist  in insulators.
}
{\begin{tabular}{@{}l lll@{}} \Hline 
\\[-1.8ex] 
 \textbf{Features}   &  Conduction electron spin current  & Magnon currents (3-dim)    &   J-W spin currents (1-dim)   \\[0.8ex]  
\hline \\[-1.8ex] 
$\bullet  $ Systems  &  Metals  &  Insulators  &   Insulators  \\
$\bullet  $ Spins &   Conduction electrons' spins    &   Localized spins   &    Localized spins      \\  
$\bullet  $ Statistics   &  Fermions     & Bosons    &     Fermions          \\
$\bullet  $ Decay length  &  A few nano-meters    &  A few centi-meters \cite{spinwave}     &   $  \setminus  $        \\[0.8ex]  
\Hline \\[-1.8ex] 
\end{tabular}}
\label{tab1}
\end{table}
\section{ Jordan-Wigner Spin Current }
\label{sec:intro}

The Mermin-Wagner theorem
states that continuous symmetries cannot be spontaneously broken at the finite temperature 
in one- and two-dimensional systems.
Therefore in such low-dimensional spin systems, the Holstein-Primakoff transformation is not useful.
In other words, because the one-dimensional XXZ model has SO(2) symmetry
that the magnon description cannot be well applied.
Thus in one-dimensional spin systems,
the spin degrees of freedom cannot be mapped efficiently into boson degrees of freedom,
but it can be well mapped into spinless fermion degrees of freedom via the J-W transformation.
Then the XY model reduces to  a free fermion system. 

\subsection{ Definition }
\label{subsec:JW}

The J-W transformation maps the spin operators of the XXZ model onto fermionic operators;
$ S_i^+    =   c_i^{\dagger }  \Pi _{j=1}^{i-1} (1-2n_j),    
 S_i^-     =    c_i  \Pi _{j=1}^{i-1} (1-2n_j),   
  S_i^z   =    n_i -   1/2,   \ \ 
  n_i      \equiv    c_i^{\dagger } c_i.$
Operators $c^{\dagger }/c $ are J-W fermion creation/annihilation operators
which satisfy the fermionic commutation relation.
The operator,  $ n     \equiv    c^{\dagger } c  $, is the number operator of the J-W fermion.
This spinless fermion carries the spin angular momentum of localized spins in one-dimensional systems. 
The z-component (i.e. $ S_i^z   =    n_i -   1/2  $)  shows that a spin down can be viewed as an empty site
and a spin up corresponds to the presence of a  fermion. 
After the J-W transformation and the canonical transformation;
$    c_{i}   \rightarrow    (-1)^{i}   c_{i}   $, 
$  {\mathcal{H}}_{\rm{XXZ}}   $ reads
\begin{eqnarray}
 {\mathcal{H}}_{\rm{XXZ}} =   -   \frac{\Gamma }{2}    \sum_{i}     ( c_i^{\dagger }  c_{i+1}  +  c_{i+1}^{\dagger }  c_{i} )  
                                      +   \Gamma       \sum_{i}   \Delta_{i}   ( n_i-\frac{1}{2} )(n_{i+1}-\frac{1}{2}  ).  
\label{eqn:c2-2}
\end{eqnarray}
Eq.(\ref{eqn:c2-2})  shows that 
only  fermions between neighbors  feel an interaction $ \Gamma \Delta  $
and J-W fermions hop between neighboring site with a hopping matrix element  $ \Gamma /2 $.
Therefore in the continuous  limit,  
$ {\mathcal{H}}_{\rm{XXZ}} $   reads   ${\mathcal{H}}_0  +  V_{\Delta }  $  
as an effective Hamiltonian\footnote{Through this paper, we take $ \hbar =1  $ for convenience.};
\begin{eqnarray}
{\mathcal{H}}_0   =   \int   dx   c^{\dagger }(x,t)    ( - \frac{1}{2m} \frac{d^2}{dx^2})   c(x,t),  \label{eqn:H0}   \label{eqn:effectivemass}  \\
  V_{\Delta }           =   \int   dx   J(x)    c^{\dagger }(x,t)   c^{\dagger }(x,t) c(x,t)  c(x,t).  
\label{eqn:Vz}
\end{eqnarray}
Here the parameter $  m  $ is the effective mass of a J-W fermion which is related to the curvature of the dispersion relation,
and
the parameter $ J $ corresponds  to $ - \Gamma \Delta  $ in the discrete model.

Moreover,   there are magnetic impurities    in  real materials.
Magnetic impurity scatterings  make the lifetime of quasiparticles $ \tau $, 
which is inversely proportional to  the imaginary part of the self-energy,  finite. 
Here it should be emphasized that 
the lifetime of quasiparticles in insulators is generally 
far longer than that of conduction electrons in metals
because the dissipation from Joule heating does not exist  in insulators;
the lifetime of conduction electrons in metals is mainly caused by 
lattice defects, nonmagnetic impurities and magnetic impurities et al. through Coulomb interactions.
Because quasiparticles in insulators have no charge degrees of freedom that the influence of impurities  is far smaller than that in metals;
magnetic impurities, at most, might cause impurity scatterings.
Furthermore in real materials, the rate of impurities such as lattice defects and nonmagnetic impurities is, in general, far larger than that of magnetic impurities.
Therefore the imaginary part of the self-energy due to magnetic impurity scatterings can be assumed to be so small that
the dispersion relation does not change (i.e. $\omega _{{{k}}} \propto k^2$), and  
we omit the vertex correction (i.e. the diffusive spin current\cite{takeuchi}).
Let us mention that in eq.(\ref{eqn:effectivemass}) we have  defined the effective mass of a J-W fermion, $m$, 
which includes  the contribution  of the self-energy due to magnetic impurity scatterings.

Thus we introduce the lifetime as a phenomenological parameter 
and assume that it is independent of temperatures. 
In order to take account of magnetic impurity scatterings,
we adopt   retarded and advanced Green's functions,
$ G^{\rm{r}}_{ {{k}}, \omega } =  [ \omega - \omega _{{{k}}} + i/ (2\tau) ]^{-1}  = (G^{\rm{a}}_{ {{k}}, \omega })^{*} $  where  $\omega _{{{k}}} = k^2/(2m)  $,
on perturbative expansion in Subsec.\ref{subsec:magnetic} and   Subsec.\ref{subsec:xxz}.
Through  these  procedure,\cite{haug}
the effects of magnetic impurity scatterings, which are the essential dynamics in real materials, are phenomenologically included into our calculation.

The density of J-W fermions , $\rho_{\rm{JW}} $, of the system is defined as the expectation value
of the number operator of J-W fermions as
\begin{equation}
\rho_{\rm{JW}}   ({x},t)   \equiv    \langle   c^{\dagger } ({x},t)   c({x},t)   \rangle.\;
\label{eqn:a2}
\end{equation}
Through Heisenberg's equation of motion,
the spin current density carried by J-W fermions, i.e. the J-W spin current,
 $ j_{\rm{JW}} $,  is defined as 
\begin{eqnarray}
    \partial_t    \rho_{\rm{JW}}      +   \partial _x   {j}_{\rm{JW}}  =  0,   \\
     {j}_{\rm{JW}}     \equiv    \frac{1}{m}     {\rm{Re}}    [ i  \langle   (\partial _{x}  c^{\dagger })   c    \rangle  ].
\label{eqn:a3}
\end{eqnarray}

\subsection{ Evaluation: magnetic field  along the quantization axis }
\label{subsec:magnetic}

In this subsection, we consider the $ J=0 $ case
and evaluate the J-W spin current generated by 
magnetic field gradients along the quantization axis.
We treat  $  V_{\rm{B}} $;
 $  V_{\rm{B}}  =  - \int   dx  B(x)   S^z  =  - \int   dx    B(x)   [c^{\dagger }(x,t)c(x,t) -   1/2]  $,
as a perturbative term to evaluate
the first-order contribution in $B$ without vertex corrections.

Through the standard procedure of the Schwinger-Keldysh (or contour-ordered) Green's function,\cite{ramer,kamenev,kita}
the Langreth method,\cite{haug,tatara,new}  the J-W spin current $j_{\rm{JW}}^{ {\rm{B}}   }  $
 is evaluated  as
\begin{eqnarray}
   j_{\rm{JW}}^{ {\rm{B}}   }&=  \frac{1}{{ 2 \pi mL^2  }}    {\rm{Im}}     \sum_{kq}    (k + {q}/{2})    {\rm{e}}^{-iqx} B_{-q}   \int    d\omega      \nonumber   \\
                      &\cdot ( G^{\rm{>}}_{k+q/2, \omega }     G^{\rm{<}}_{k-q/2, \omega }   -  G^{\rm{t}}_{k+q/2, \omega }    G^{\rm{t}}_{k-q/2, \omega }    )  +O(B^2). 
\label{eqn:n7}
\end{eqnarray}                      
The variable  $ L $ is the chain length and 
the variables  $   G^{\rm{t}}$, $  G^{\rm{>}} $, and  $  G^{\rm{<}} $ 
are the time-ordered, greater, and lesser Green's functions, respectively.  
Fermionic greater and lesser Green's functions are defined as
$    G^{>} (t_1, t_2)  \equiv  -i  \langle  c(t_1) c^{\dagger } (t_2)  \rangle,   G^{<} (t_1, t_2) \equiv  i  \langle  c^{\dagger }(t_2)   c(t_1)  \rangle          $.
They satisfy the relations:
   $ G^{\rm{t}}_{k, \omega }   =    G^{\rm{a}}_{k, \omega }     +   G^{\rm{>}} _{k, \omega } $,
   $G^{\rm{>}}_{k, \omega }   -   G^{\rm{<}}_{k, \omega }    =   G^{\rm{r}}_{k, \omega }   -  G^{\rm{a}}_{k, \omega }$.
Thus the current is rewritten as
\begin{eqnarray}
 j_{\rm{JW}} ^{ {\rm{B}}  }&= -\frac{ 1  }{ 2 \pi mL^2  }      {\rm{Im}}     \sum_{kq}    (k + \frac{q}{2})     {\rm{e}}^{-iqx}    B_{-q}  \int  d\omega    \nonumber  \\    
                                  &\cdot ( G^{\rm{a}}_{k+q/2, \omega }   G^{\rm{a}}_{k-q/2, \omega }     +  G^{\rm{a}}_{k+q/2, \omega }   G^{\rm{>}}_{k-q/2, \omega }   
                                                                         + G^{\rm{>}}_{k+q/2, \omega }   G^{\rm{r}}_{k-q/2, \omega }    ).    
\label{eqn:b4}
\end{eqnarray}
The retarded Green's function, $ G^{\rm{r}}$, is given by 
  $ G^{\rm{r}}_{ {{k}}, \omega } =  [ \omega - \omega _{{{k}}} + i/ (2\tau) ]^{-1}  =  (G^{\rm{a}}_{ {{k}}, \omega })^{\ast } $
and  $\tau$ is the lifetime of a J-W fermion.
The energy $\omega _{{k}}$ is $ \omega _{{k}} = Fk^2 $,    $ F    \equiv   (2m)^{-1}   $.
Then the function $  G^{\rm{r}} $ reads
$ G^{\rm{r}}_{k-q/2, \omega }     =      G^{\rm{r}}_{k, \omega }    +     ( G^{\rm{r}}_{k, \omega } )^2   (  - Fkq  +  {Fq^2}/{4}  )  +O (  (kq)^2).$
This  approximation   demands that 
the magnetic field varies slowly and moderately in space (compared to the  J-W fermion  mean-free path),  
and  the value of  $ \partial _x B $ is a constant.
Then eq.(\ref{eqn:b4}) reads
\begin{eqnarray}
 j_{\rm{JW}} ^{ {\rm{B}}  }&= -\frac{ 4 F \tau  }{   L^2  }      {\rm{Im}}     \sum_{kq}    (k + \frac{q}{2})     {\rm{e}}^{-iqx}   i   B_{-q}      \nonumber  \\    
                                  &\cdot  \Big\{     [  4{\tau} (1-n_{k})   +i  \beta   {\rm{e}}^{\beta (Fk^2-\mu)}   {n_{k}}^2    ]Fkq      \nonumber  \\       
                                            &-    [  4{\tau} (1-n_{k})   -i  \beta   {\rm{e}}^{\beta (Fk^2-\mu)}   {n_{k}}^2    ]Fkq             \Big\} ,                                                 
\label{eqn:n2}
\end{eqnarray}
where  $ n_k $ is the Fermi distribution function, $ \mu $ is the chemical potential, $\beta \equiv 1/ (k_{\rm{B}} T) $ and
$  k_{\rm{B}} $   is the Boltzmann constant. 
We have used the approximation;
$   n_{k+q/2} \approx  n_{k} -\beta F     {\rm{e}}^{\beta (Fk^2-\mu)}    n_{k}^2  kq $.
By the Sommerfeld expansion ($   T \ll  T_{\rm{F}},    T_{\rm{F}} $ ; the  Fermi temperature\footnote{
The Fermi temperature of this system(insulator) is estimated as a few hundred Kelvin. }),  the chemical potential   is evaluated  as
$  \mu(T)  =   \epsilon _{\rm{F}}  +    {(\pi   k_{\rm{B}}T)^2}/({6   \epsilon _{\rm{F}}})    +  O(T^4), $
where $  \epsilon _{\rm{F}}  $  is  the Fermi energy.

Finally, the J-W spin current generated by the magnetic field gradient along the quantization axis 
 in the finite temperature  near zero Kelvin     reads
\begin{eqnarray}
    j_{\rm{JW}} ^{ {\rm{B}}  }  =   \frac{ 4 \sqrt{F} \tau }{  \pi   L      }  ( {\rm{ln}}2 + \frac{{\pi}^2}{24} )   
                               \sqrt{  \epsilon _{\rm{F}}  + \frac{(\pi   k_{\rm{B}}  T)^2  }{   6  \epsilon _{\rm{F}}}   }  \partial _x B.
\label{eqn:b14}
\end{eqnarray}

\subsection{ Evaluation: two-particle interaction }
\label{subsec:xxz}

In this subsection, we  treat the $ B=0 $ case
and evaluate the J-W spin current generated by 
the two-particle interaction gradient.
The two-particle interaction arises from the anisotropic exchange interaction  $  V_{\Delta }    $;
 $  V_{\Delta } = \int dx J(x)  c^{\dagger }(x,t)   c^{\dagger }(x,t) c(x,t)  c(x,t),   $
and we treat it as a perturbative term to evaluate
the first-order contribution in $J$ without vertex corrections.

Through the same procedure with  the last subsection, 
the J-W spin current generated by the two-particle interaction  $V_{\Delta }  $  is evaluated as  
\begin{eqnarray}
 j_{\rm{JW}} ^{\Delta }&=    -\frac{ F  }{  {\pi}^2 L^3  }      {\rm{Im}}     \sum_{kq}    (k + \frac{q}{2})     {\rm{e}}^{-iqx}    J_{-q} i\sum_{\tilde p}  \int  d{\omega}'    G^{\rm{<}}_{\tilde p, {\omega}' }   \int  d\omega   \nonumber  \\    
                                  &\cdot (     G^{\rm{a}}_{k+q/2, \omega }   G^{\rm{a}}_{k-q/2, \omega }     
                                                 +   G^{\rm{a}}_{k+q/2, \omega }   G^{\rm{>}}_{k-q/2, \omega }     
                                                 + G^{\rm{>}}_{k+q/2, \omega }   G^{\rm{a}}_{k-q/2, \omega }     \nonumber  \\ 
                                                 &+ G^{\rm{>}}_{k+q/2, \omega }   G^{\rm{>}}_{k-q/2, \omega }    
                                                 -  G^{\rm{>}}_{k+q/2, \omega }   G^{\rm{<}}_{k-q/2, \omega }    )                   +O(J^2).
  \label{eqn:n3}                                 
\end{eqnarray}
We have used the relation that  the number operator, $    \langle c^{\dagger } (t) c(t)   \rangle $,
is the same as the equal-time lesser Green's function\cite{haug,new};
$    \langle c^{\dagger } (t) c(t)   \rangle  = -iG^{<} (t,t)     $. 
 Because  each  Green's  function is not independent; $ G^{\rm{r}} - G^{\rm{a}} =  G^{>} - G^{<}  $,
 that eq.(\ref{eqn:n3}) is rewritten as
\begin{eqnarray}
 j_{\rm{JW}} ^{\Delta }&=    -\frac{ F  }{  {\pi}^2 L^3  }      {\rm{Im}}     \sum_{kq}    (k + \frac{q}{2})     {\rm{e}}^{-iqx}    J_{-q} i\sum_{\tilde p}  \int  d{\omega}'    G^{\rm{<}}_{\tilde p, {\omega}' }   \int  d\omega   \nonumber  \\    
                                  &\cdot (     G^{\rm{a}}_{k+q/2, \omega }   G^{\rm{a}}_{k-q/2, \omega }     
                                                 +   G^{\rm{a}}_{k+q/2, \omega }   G^{\rm{>}}_{k-q/2, \omega }     
                                                 + G^{\rm{>}}_{k+q/2, \omega }   G^{\rm{r}}_{k-q/2, \omega }). 
  \label{eqn:n99}                                 
\end{eqnarray}
Also in this case,
we  assume that the two-particle interaction varies slowly and moderately in space (compared to the J-W fermion mean-free path),  
and  the value of  $ \partial _x J $ is a constant.
Therefore the retarded green's function can be approximated;
$ G^{\rm{r}}_{k-q/2, \omega }     =      G^{\rm{r}}_{k, \omega }    +     ( G^{\rm{r}}_{k, \omega } )^2   (  - Fkq  +  {Fq^2}/{4}  )  +O (  (kq)^2).$
Then eq.(\ref{eqn:n99}) reads
\begin{eqnarray}
 j_{\rm{JW}} ^{\Delta }   =    -   \frac{ 16  F \tau }{  L^3 }   {\rm{Re}} \sum_{k} \beta F   {\rm{e}}^{\beta (Fk^2-\mu)} k^2 n_{k}^2     
                                 \sum_{\tilde p}     n_{\tilde p}   \partial _x J. 
  \label{eqn:n77}                                 
\end{eqnarray}
Finally, the J-W spin current generated by the two-particle interaction    in the finite temperature   near zero Kelvin   reads
\begin{eqnarray}
     j_{\rm{JW}}^{\Delta }=  - \frac{ 4 \tau }{  {\pi}^2   L  }    (    {\rm{ln} } 2    +  \frac{{\pi}^2}{24}  )  
                                     \Bigg\{    2    \Big[    \epsilon _{\rm{F}}    +    \frac{(   \pi    k_{\rm{B}}T    )^2}{6    \epsilon _{\rm{F}}}     \Big]  
                         -  \frac{(   \pi     k_{\rm{B}} T)^2}{   12 \Big[ \epsilon _{\rm{F}}  + \frac{(  \pi   k_{\rm{B}}T)^2}{6   \epsilon _{ \rm{F}}}  \Big]  }  \Bigg\}    \partial _x J.                                                 
\label{eqn:c8}
\end{eqnarray}

\section{Magnon Current }
\label{sec:mag}

According to the Mermin-Wagner theorem,
three-dimensional spin systems are the most suitable set-ups for the study of magnon transports.
Therefore in order to clarify the thermal properties of magnon currents,
we focus on three dimensional ferromagnetic insulators at the low-temperature near zero Kelvin,
which has no conduction electrons.
The spin degree of freedom reduces to the magnon degree of freedom 
via the Holstein-Primakoff transformation;
the transverse components of the exchange interactions
give rise to hopping of the magnons, while the longitudinal
component give rise to the interaction.

External magnetic fields, of course,   generate magnon currents also in this case;
the magnetic field along the quantization axis is equivalent to a chemical potential for magnons as well as J-W fermions,
and the gradient generates magnon currents.
Moreover time-dependent transverse magnetic fields cause quantum fluctuations and  generate
a magnon current through   pumping effects.\cite{knpump}
Here, however, we consider three dimensional ferromagnetic insulators  without any external magnetic fields.

\subsection{ Definition }
\label{subsec:def}

The low-energy effective Hamiltonian\cite{oshikawa} for magnons in three-dimensional  insulators
reads  $ {\cal{H}}_{\rm{m}}  = {\cal{H}}_{\rm{0(m)}} + V_{\rm{m}} $; 
\begin{equation}
{{\cal{H}}}_{\rm{0(m)}}   =  \int d^3x a^{\dagger }(\mathbf{x},t) \Big(-\frac{ {\mathbf{\nabla }}^2 }{2m_{\rm{mag}}}  \Big)  a(\mathbf{x},t),
\end{equation}
\begin{equation}
        V_{\rm{m}}    = \int d^3x J_{(\rm{m})} (\mathbf{x})  a^{\dagger }(\mathbf{x},t) a^{\dagger }(\mathbf{x},t)  a(\mathbf{x},t) a(\mathbf{x},t). 
\end{equation}
Here $m_{\rm{mag}} $ is the effective mass of a magnon,
and $ V_{\rm{m}}  $ represents the magnon-magnon interaction.
Operators $a^{\dagger }/a $ are magnon creation/annihilation operators
which satisfy the bosonic commutation relation.
In the  low-temperature,
only the two-particle interaction $ J_{(\rm{m})}  $ is important.\cite{model1,model2}

The  magnetic impurity scattering is, of course,  an inevitable dynamics in real materials
even when the system is under low-temperature.
It  makes the lifetime of magnons,  $ \tau_{\rm{m}} $,  finite. 
Therefore  by the same procedure with Subsec.\ref{subsec:JW},
we phenomenologically include the effects of magnetic impurity scatterings, which are the inevitable dynamics in real materials, into our calculation.
Also in this case,
we have assumed that  the imaginary part of the self-energy due to magnetic impurity scatterings is so small that
the dispersion relation does not change,  and  
we omit the vertex correction (i.e. the diffusive spin current). 
Furthermore, we assume that the lifetime of magnons is  independent of temperatures.  

The magnon density, $\rho_{\rm{m}} $,  of the system is defined as the expectation value
of the number operator of magnons
\begin{equation}
\rho_{\rm{m}} (\mathbf{x},t)\equiv \langle a^{\dagger }(\mathbf{x},t)a(\mathbf{x},t)\rangle.\;
\label{eqn:A2}
\end{equation}
Through Heisenberg's equation of motion,
the magnon current (density), $\mathbf{ j_{\rm{m}}} $,  is defined as 
\begin{eqnarray}
\frac{\partial  \rho_{\rm{m}} }{\partial  t}  + \mathbf{\nabla} \cdot \mathbf{ j_{\rm{m}}} =0.
\label{eqn:A3}
\end{eqnarray}
Then the magnon current reads
\begin{eqnarray}
j_{\rm{m}}^{\nu } (\mathbf{x},t) = \frac{1}{m_{\rm{mag}}}   {\rm{Re}}  [i \langle (\partial_\nu a^{\dagger }(\mathbf{x},t))a(\mathbf{x},t) \rangle ] ,\;
\label{eqn:A4}
\end{eqnarray}
where $ \nu $ is a direction for a magnon current to flow ($\nu = x,y,z  $).

\subsection{ Evaluation }
\label{subsec:eva}

From now on, we treat the magnon-magnon interaction term,
$  V_{\rm{m}}   $,
as a perturbative one
to evaluate
the first-order contribution in $ J_{(\rm{m})}  $ without vertex corrections.
Through the same procedure with the last section, the Langreth method,\cite{haug,tatara,new} 
the magnon current 
 is evaluated as
\begin{eqnarray}
 j_{\rm{m}}^{\nu }&= - \frac{4}{ m_{\rm{mag}}  }  (\frac{2\pi}{V})^3      {\rm{Re}}   
        \sum_{\mathbf{k,q} } (k_{\nu} + \frac{q_{\nu}}{2} )        {\rm{e}}^{ -i  \mathbf{q} \cdot {\mathbf{x}} }    {J }_{(\rm{m}) -\mathbf{q}}      
         \sum_{\mathbf{\tilde p}}    \int 
        \frac{d{\omega}'}{2\pi }  \tilde G^{\rm{<}}_{ {\mathbf{\tilde p}}, {\omega}'}    \int  \frac{d\omega }{2\pi }         \nonumber  \\
      &\cdot      ( \tilde G^{\rm{a}}_{ {\mathbf{k}}  + {\mathbf{q}}/2, \omega }          \tilde G^{\rm{a}}_{ {\mathbf{k}}  - {\mathbf{q}}/2, \omega } 
               +\tilde G^{\rm{a}}_{ {\mathbf{k}}  + {\mathbf{q}}/2, \omega }          \tilde G^{\rm{>}}_{ {\mathbf{k}}  - {\mathbf{q}}/2, \omega }  
             +\tilde G^{\rm{>}}_{ {\mathbf{k}}  + {\mathbf{q}}/2, \omega }          \tilde G^{\rm{r}}_{ {\mathbf{k}}  - {\mathbf{q}}/2, \omega } ) + O(J_{(\rm{m})}^2).     
\label{eqn:koreda}                              
\end{eqnarray}
Here $V$ is a volume of the system.
Bosonic greater and lesser Green's functions are defined as
$    \tilde G^{>} (t_1, t_2)  \equiv  -i  \langle  a(t_1) a^{\dagger } (t_2)  \rangle,   \tilde G^{<} (t_1, t_2) \equiv  -i  \langle  a^{\dagger }(t_2)   a(t_1)  \rangle          $;
the sign of the lesser green's function is different from the fermionic case.
The retarded Green's function is  $\tilde G^{\rm{r}}_{ {\mathbf{k}}, \omega } =  [\omega - \tilde \omega _{{\mathbf{k}}} + i/( 2\tau_{\rm{m}})]^{-1}  =  (\tilde G^{\rm{a}}_{ {\mathbf{k}}, \omega } )^{\ast } $,
where the energy $\tilde \omega _{\mathbf{k}}$ is
$ \tilde \omega _{\mathbf{k}} = Dk^2 $, $ D\equiv 1 /(2m_{\rm{mag}}) $
and 
$\tau_{\rm{m}}$ is the lifetime of magnons.
The lifetime  represents the damping of spins
($\tau_{\rm{m}} $ is inversely
proportional to the Gilbert damping parameter, $\alpha $).
Then the retarded Green's function satisfies 
     $ \tilde G^{\rm{r}}_{ {\mathbf{k}} + {\mathbf{q}}/2 , \omega   }    
=   \tilde G^{\rm{r}}_{ {\mathbf{k}} , \omega   }  +  (\tilde G^{\rm{r}}_{ {\mathbf{k}} , \omega   })^2  ( D {\mathbf{k}}\cdot {\mathbf{q}} +{q^2}/{4} )   
     + O  ( ({ \mathbf{k}}\cdot {\mathbf{q}})^2  ).$
This approximation demands that 
the magnon-magnon interaction  varies slowly and moderately (compared to the magnon mean-free path), 
and  the value of $ \partial _{\nu  }J_{(\rm{m})}  $ is a constant.

Thus  eq.(\ref{eqn:koreda})  is rewritten as
\begin{eqnarray}
   j_{\rm{m}}^{\nu }   &=  \frac{8  D^2  \tau_{\rm{m}}  }{   V^2}       {\rm{Re}}   
        \sum_{\mathbf{k,q} } (k_{\nu} + \frac{q_{\nu}}{2} )        {\rm{e}}^{ -i  \mathbf{q} \cdot {\mathbf{x}} }    {J }_{(\rm{m}) -\mathbf{q}}      
             \    {i }    \    \zeta (\frac{3}{2})  \Gamma (\frac{3}{2})  (\frac{k_{\rm{B}}T}{D})^{3/2}    \nonumber  \\
      &\cdot    \Big\{    [  \beta   ( {\rm{e}}^{-\beta Dk^2} +2{\rm{e}}^{-2\beta Dk^2}   ) + 4i {\tau_{\rm{m}}}  (1+\tilde n_{\mathbf{k}})  ]   {\mathbf{k}} \cdot  {\mathbf{q}}         \nonumber  \\
                       &+  [   \beta   ( {\rm{e}}^{-\beta Dk^2} +2{\rm{e}}^{-2\beta Dk^2}   ) - 4i {\tau_{\rm{m}}}  (1+\tilde n_{\mathbf{k}})  ]   {\mathbf{k}} \cdot  {\mathbf{q}}         \Big\}  \\
                             &=  \frac{ 16 \tau_{\rm{m}}  }{3    V^2 }  \zeta (\frac{3}{2})  \Gamma  (\frac{3}{2})   \sqrt{ D k_{\rm{B}} T }     
                             \  {\rm{Re}}   \sum_{\mathbf{k} } k^2    (  {\rm{e}}^{-\beta Dk^2}  + 2 {\rm{e}}^{-2\beta Dk^2}  )   
                        \sum_{\mathbf{q}}  iq_{\nu}    {\rm{e}}^{ -i  \mathbf{q} \cdot {\mathbf{x}} }       {J }_{{(\rm{m})}  -\mathbf{q}}. 
\label{eqn:B6}                              
\end{eqnarray}
Here $ \tilde n_{\mathbf{k}} $ is the Bose distribution function,
$ \zeta  $ is the Riemann zeta function,
$ \zeta ( 3/2 ) =2.612  $, 
and $ \Gamma $ is the Euler gamma function,
$  \Gamma (3/2) = \sqrt{\pi}  /2$.
Because we consider the the  low-temperature  regime  near zero Kelvin,
we have  approximated as 
$     \tilde n_{\mathbf{k}+\mathbf{q} /2 }      \approx  \tilde n_{\mathbf{k}}  -\beta D    {\rm{e}}^{-\beta Dk^2}   (1- {\rm{e}}^{-\beta Dk^2} )^{-2}   {\mathbf{k}} \cdot  {\mathbf{q}} 
                                                               \approx    \tilde n_{\mathbf{k}}  -\beta D  ({\rm{e}}^{-\beta Dk^2}  +2 {\rm{e}}^{-2\beta Dk^2}  )   {\mathbf{k}} \cdot  {\mathbf{q}}       $.
Moreover we have used the isotropy condition in the direction for $ {\mathbf{k}} $; 
$  \sum_{\mathbf{k} }  k_{\nu}   { \mathbf{k}}\cdot {\mathbf{q}}  =  \sum_{\mathbf{k} }  k^2 q_{\nu} /3 $.

Finally   in the low-temperature  near zero Kelvin,
the magnon current generated by the magnon-magnon interaction gradient reads 
\begin{eqnarray}
       j_{\rm{m}}^{\nu }= - \frac{ D  \tau_{\rm{m}} }{  V }   (1+\frac{1}{2^{3/2}} )   \zeta (\frac{3}{2})    \Gamma  (\frac{3}{2})   
                               \Big(\frac{ k_{\rm{B}} T }{  D \sqrt{\pi}    } \Big)^{3}    \partial _{\nu}   J_{ {(\rm{m})}  }.
\label{eqn:B4}
\end{eqnarray}
\begin{table}[ph]   
\tbl{The temperature dependence of  J-W spin currents and magnon ones
generated by the magnetic field gradient
and  by the two-particle interaction gradient.  
The constant $ \gamma   $ is defined as 
$    \gamma    \equiv   6 [\epsilon _{\rm{F}} / (\pi k_{\rm{B}} ) ]^2     $.
J-W spin currents  exhibit the stronger  dependence on temperatures than  magnon ones
in the finite temperature near zero Kelvin (i.e. $  0 \not=   T<1 \    $[K]).
}
{\begin{tabular}{@{}lll@{}} \Hline 
\\[-1.8ex] 
 \textbf{Temperature dependence}  &   J-W spin currents (1-dim)  & Magnon currents (3-dim)  \\[0.8ex]  
\hline \\[-1.8ex] 
$\bullet  $ Magnetic field;  $ \partial _{\nu} B  $     &  $       \propto    \sqrt{\gamma  +T^2}  $    &        $ \propto  T^{3/2}    $  (Ref.\refcite{spinwave}, \refcite{taguchi})  \\ 
$\bullet  $ Two-particle interaction;  $  \partial _{x} J  $,  ($  \partial _{\nu} J_{(\rm{m})}  $) &  $    \propto   [ \gamma  +T^2 - T^2/(4+4T^2/\gamma  )  ]$  &       $ \propto T^3$    \\[0.8ex] 
\Hline \\[-1.8ex] 
\end{tabular}}
\label{tab2}
\end{table}
\section{Temperature Dependence near Zero Kelvin  }
\label{sec:sc}

The Mermin-Wagner theorem gives crucial differences
between low-dimensional spin systems and three-dimensional ones.
It prohibits the spontaneous breaking of the continuous symmetry of the low-dimensional XXZ model, i.e. SO(2) symmetry, in the finite temperature.
In other words, in low-dimensional systems quantum fluctuations are so strong as to destroy magnetic orders.
Then the thermal properties of each spin current in insulators, 
the J-W spin current and the magnon one, 
are drastically different from each other in the finite temperature near zero Kelvin (i.e. $ 0  \not= T<1 \   $ [K]);
as shown in  Table  \ref{tab2},
the J-W spin current  exhibits the stronger  dependence on temperatures than the magnon current, i.e.
the exponents of temperatures in J-W spin currents are smaller than 
those in magnon currents.
Thus, the influence of quantum and thermal fluctuations  in one-dimensional insulators 
 is larger than that in three-dimensional ones. 
The Table \ref{tab2}   also shows that 
the current generated by the two-particle interaction gradient
shows the weaker dependence on temperatures
than the one by the magnetic field gradient along the quantization axis.

\section{ Summary and Discussion }
\label{sec:sum}

We have theoretically studied the temperature dependence of the spin current in one- and three-dimensional insulators.
In one-dimensional insulators,
the spin current is carried by Jordan-Wigner fermions.  
In this system,
quantum fluctuations are  strong and 
the J-W spin current shows the stronger dependence on temperatures than the magnon one
in the finite temperature near zero Kelvin.

Experimentally though the direct measurement of spin currents is impossible at this stage,
the spin currents we have discussed 
would be identified by observing the temperature dependence  of the spin currents through inverse spin-Hall effects.
Theoretically, the microscopic calculation of the lifetime caused by magnetic impurity scatterings and
also by phonon ones et al., 
and the diffusive spin current with vertex corrections
is an significant theoretical issue.

\section*{Acknowledgements}

The author would like to thank K.Totsuka, G.Tatara, S.Fujimoto and  M.Oshikawa.
He is also grateful to Y.Korai  and K.Taguchi for invaluable discussion
and T.Kimura for reading of the manuscript and useful comments.
We  wish  to  acknowledge  the G.Tatara-laboratory members in Tokyo Metropolitan University for warm hospitality during his stay.

This work was supported by the Grant-in-Aid for the Global COE Program
 "The Next Generation of Physics, Spun from Universality and Emergence" 
from the Ministry of Education, Culture, Sports, Science and Technology (MEXT) of Japan.

\section*{References}



\end{document}